\documentclass[final,5p,times,twocolumn]{elsarticle}

\usepackage{amssymb}
\usepackage{amsmath}
\usepackage{url}
\usepackage{framed}
\usepackage{booktabs}
\usepackage{longtable}
\usepackage{multirow}

\usepackage{enumitem}

\usepackage{xcolor}
\usepackage{tabularx}
\usepackage{rotating}

\journal{Journal of Systems and Software}

\begin{document}

\begin{frontmatter}


\title{Advancing Digital Government through Open Source Indicators}

\author[1]{Johan Linåker}\corref{cor1}
\ead{johan.linaker@ri.se}
\ead[url]{https://orcid.org/0000-0001-9851-1404}

\author[1]{Sachiko Muto}
\ead{sachiko.muto@ri.se}

\cortext[cor1]{Corresponding author.}

\address[1]{RISE Research Institutes of Sweden AB, Lund, Sweden}




\begin{abstract}
\textbf{Context}:
Open Source Software (OSS) is a vital public good that underpins nearly all modern software stacks. Its licensing enables reuse and collaboration, fueling innovation, competitiveness, and digital sovereignty. European and global policy frameworks recognise OSS as a key enabler of public sector transformation, advancing interoperability, transparency, and technological independence. However, systematic awareness and measurement of governmental OSS adoption remain limited.
\textbf{Objectives}:
This study contributes to digital government maturity indexes by analysing how policies and institutional actions leverage OSS for software reuse and collaborative development across 16 digitally mature countries. It proposes potential indicators for inclusion in such indexes by examining OSS policy goals, key actors, and support mechanisms.
\textbf{Methods}:
A qualitative approach combines desk research of policy documents with semi-structured interviews of government representatives, producing detailed country reports. Cross-analysis focuses on policy rationales, promotion strategies, and implementation support.
\textbf{Results}:
OSS reuse policies are widespread, targeting both inbound adoption and outbound sharing, and are often led by central public sector organisations. Policy objectives emphasise interoperability, sovereignty, transparency, cost efficiency, and security—framed as both challenge and strength. Implementation is supported by Open Source Program Offices (OSPOs) that facilitate capacity building, resource pooling, and sustainable governance. Thirteen indicator areas are proposed, spanning policy design, incentives, implementation, and support.
\textbf{Conclusions}:
OSS is a strategic enabler of digital transformation. Realising its full value requires coherent policy frameworks, institutional capacity, and integration of OSS indicators within global maturity assessments to enhance transparency, interoperability, and sovereignty.
\end{abstract}




\end{frontmatter}




\section{Introduction}
\label{sec:introduction}

Open Source Software (OSS) constitutes a public good, estimated to be present in 96 per cent of today’s software stacks~\cite{synopsis2025survey}. The OSS licensing regime allows the software to be freely reused, inspected, modified, and redistributed, and by extension enables a collaborative form for development shown to have a substantive impact on both the GDP~\cite{blind2021impact}, and increased competitiveness and growth of national tech industries~\cite{nagle2019government}. A study brings light on the value generated both from a supply and demand side perspective, with figures between \$1.22-6.22 billion, and \$2.59-13.18 trillion respectively~\cite{hoffmann2024value}. The potential of OSS as an instrument for software reuse and collaborative development is accordingly significant and a pivotal tool for the digital transformation of the public sector. 

At the European level, the importance has been stressed through successive Ministerial Declarations~\cite{ec2017tallin, ec2020berlin, ec2022strasbourg}, as well as in overarching visions~\cite{ec2021digitalcompass}, strategies~\cite{ec2020ossStrategy}, and legislative frameworks~\cite{ec2017interoperability}. Interoperability among digital infrastructure and service components, sovereignty in technical sourcing and design decisions, and transparency into data collection and use are some of the more prominent policy goals being reiterated. 

Also, in other parts of the world, such as the US and China, OSS has gained a prominent role in the long-term strategic programmes for digital transformation, as well as an awareness of the need to keep pace with technology development, attracting a skilled workforce, and proactively manage the increased threats of cybersecurity attacks. The importance of OSS as a carrier of innovation is further stressed in the context of developing countries and in addressing the Sustainable Development Goals set to be achieved by 2030. Despite apparent awareness on the international scale, there is limited attention to and follow-up on overall progress at both the international and national levels of government.

International rankings have shaped government behaviour since the 1970s~\cite{merry2011measuring}, driving reforms and policy shifts as nations vie for legitimacy and status~\cite{kelley2015politics, broome2018bad}. OSS is acknowledged in prominent indexes—such as those by DIAL~\cite{dial2024principles}, OECD~\cite{oecd2023index}, and WIPO’s Global Innovation Index~\cite{wipo2024index}—that highlight its roles in interoperability and innovation. Yet, OSS indicators remain narrow and offer little actionable guidance to governments, leaving many digital transformation efforts at risk of missing major opportunities if OSS is overlooked as a strategic policy tool.

In this work, we aim to provide input to these indices through an exploratory and cross-comparative analysis of the policy and to support actions to leverage OSS as an instrument for software reuse and collaborative development among national governments. We specifically examine a sample of the 16 countries identified as digitally mature, using four different digital maturity indexes. These countries are surveyed through a multiple-case study in terms of:

\begin{itemize}
    \item RQ1: How are government policies considering or promoting OSS in terms of software reuse and collaborative development? Who are the actors involved in the policy definition and enforcement?
    \item RQ2: What are the policy goals and rationale (e.g., interoperability, sovereignty, and transparency) stated in the policies for promoting and enabling software reuse and collaborative development through OSS?
    \item RQ3: How are subjects for the policies supported and enabled in terms of implementing the policy directives? 
\end{itemize}

Desk research of online material is synthesised, complemented, and validated through semi-structured interviews with government representatives from each country. Country reports are compiled for each country and cross-analysed for each research question. The country reports are presented in a recently published report~\cite{linaaker2024reuse} 
commissioned by the Danish Agency for Digital Government (Digitaliseringsstyrelsen) and Local Government Denmark (KL), aimed at providing input on how Danish Public Sector Organisations (PSOs) can specifically become better at reaping benefits by reusing existing software and creating value by developing software in a way that it can be reused. This study extends the report by focusing on how the digital maturity indexes may be tailored to account for the specific features observed among the surveyed countries related to enabling software reuse through OSS.

Specifically, we find that most countries have implemented policies on software reuse through OSS, covering both inbound acquisition and outbound sharing, with central PSOs typically responsible for their governance. While the primary focus is on public-sector use, some policies also aim to stimulate OSS uptake in the broader technology sector. Policy goals commonly reference interoperability, digital sovereignty, transparency, and cost efficiency, alongside economic arguments, with varying emphasis across national contexts. Security is framed both as a risk—due to potential exposure of vulnerabilities—and as a strength, thanks to OSS’s transparency and collaborative oversight. Policy implementation is supported by a growing ecosystem of initiatives, notably Open Source Program Offices (OSPOs) at national, institutional, and local levels, which build capacity, enable resource pooling, and sustain joint projects. Success cases often originate from single-entity initiatives that evolve into collaboratively maintained projects, with software catalogues serving as important tools for promoting and enabling reuse.

Based on our findings, we synthesise and propose a comprehensive list of indicators per the two categories: policy incentives and design, and policy implementation and support. The list provides a foundation for research and practice to narrow down and use as inputs to indexes for measuring how governments are leveraging OSS as a tool in their digital transformation. In sum, the paper makes the following contributions:
\begin{itemize}
    \item First, we provide a cross-comparative analysis of OSS-related policies and support actions across 16 digitally mature countries, integrating perspectives on policy design and implementation. While prior work has surveyed OSS policies across regions or countries, these studies typically focus on the presence of policies or specific instruments. In contrast, we provide a more holistic view that also accounts for how such policies are operationalised in practice.

    \item Second, we explicitly distinguish between inbound (acquisition) and outbound (release) OSS policies. While this distinction is implicit in parts of the literature, we make it explicit and use it to structure the analysis, enabling a more fine-grained understanding of policy direction, scope, and intervention types.

    \item Third, we synthesise a framework capturing key dimensions of OSS policy and support, including policy types, actors, and enabling mechanisms such as OSPOs, guidelines, and software catalogues, consolidating insights from a fragmented body of work.

    \item Finally, we derive and propose a set of empirically grounded indicators for assessing OSS enablement in digital government. Existing digital government indexes include only limited and often narrow indicators related to OSS; our work provides a more comprehensive foundation for benchmarking and future index development.
\end{itemize}

The paper is structured as follows: In section~\ref{sec:background}, we provide background and related work on OSS policy and implementation. In section~\ref{sec:researchdesign}, we describe the research design and methodology used in the study. Section~\ref{sec:findings} presents the findings of the study, followed by a discussion and a presentation of a synthesised set of potential indicators for digital government indexes and governments to consider. Section~\ref{sec:conclusions} concludes the paper.

\section{Background and related work}
\label{sec:background}
Below, we elaborate on the background and related work on OSS policies in general, the motives underpinning these policies, and the challenges and actions related to their implementation. The elaboration is neither systematic nor complete, but rather provides an introduction and connects the technical and policy perspectives of OSS, which this study bridges. Purposive sampling of the literature has been used accordingly~\cite{patton2014qualitative}, using both keyword searches and snowballing in scientific databases.

\subsection{Open Source Software policies}
Policies supporting the adoption and collaboration with OSS can take many forms and have been surveyed on several occasions over the years. Bouras et al.~\cite{bouras2014policy} qualitatively survey several policies in the European context, highlighting policies on the local and regional, national, and European levels of government. Blind et al.~\cite{blind2021impact} also surveyed the OSS policies among EU countries, while also investigating Brazil, China, India, Japan, South Korea and the USA to contrast results. In more recent works, Thévenet~\cite{thevenet2024progress} surveyed EU countries as part of the EU Commission's Open Source Observatory project.

From a more quantitative perspective, the Centre for Strategic \& International Studies (CSIS) presented a mapping of 669 OSS policies across 69 countries, reporting a clear increase compared to its 2010 mapping~\cite {lostri2023gov}. The policies most commonly take the form of parliamentary bills, directives and regulations, strategies (general or OSS-specific), or more informal expressions of support for OSS. The policies typically focus on Research and Development (R\&D) and procurement, with the latter evenly split between advisory and mandatory requirements for OSS adoption. The report notes an increasing trend for the former since 2013. 

As highlighted above, gaining the support and endorsement of policymakers is critical for public sector entities~\cite{cassell2008governments, silic2017open}. The way, however, in which such support and endorsement is engraved and formulated differs. Van Loon and Toshkov~\cite{van2015adopting} report that national legislation prescribing non-mandatory OSS adoption has little effect at the local level when buy-in from local leadership is limited. An earlier report from the Finnish context showed the opposite: a general interest in OSS adoption among municipalities, while interest and support from the central government were limited~\cite{valimaki2005empirical}. 

An example of a strict approach can be found in Venezuela’s Decree 3,390 (2004), which mandates that \textit{``all systems of the government must adopt FLOSS, and that all public offices must begin a progressive and gradual adoption of FLOSS''}~\cite{maldonado2010process}. Exceptions were only permitted when no OSS alternative existed. A similar policy was under development in Peru but was later stopped – reportedly due to intense lobbying from technology incumbents~\cite{chan2017coding}. Other countries, such as France, have been reported to be more successful in implementing similar policies favouring OSS in procurement and adoption processes~\cite{nagle2023government}.  

Blind et al.~\cite{blind2021impact} report on a distinction between how policies either focus on proposing a preferential treatment to OSS in the procurement process, or on growing the skills and capacity for adopting and collaborating on OSS. While the former may be regulated through policies with varying levels of prescriptiveness, the latter is typically embodied through strategy documents \textit{``prescribing or guiding the use of Open Source within the organisation itself or guidelines on the re-use of OSS within the public sector of the jurisdiction''}~\cite{blind2021impact}.

\subsection{Policy motivations}
The motivation underpinning OSS policy varies depending on contextual factors and needs. CISA’s investigation shows that modernisation of digital infrastructure and services is a main driver in its survey of 669 OSS policies, followed by providing support for the national industry and potential cost savings~\cite{lostri2023gov}. Increased transparency, security and digital sovereignty were also highlighted. Findings overlap with the case study research by Blind et al.~\cite{blind2021impact}, highlighting both economic, technical, and political concerns that serve as motivations for policies. Below, we further nuance these areas.

\subsubsection{Economic concerns}
Considering economic concerns specifically, these may be further nuanced into potential cost savings (e.g., related to production, maintenance, and lock-in effects) and the stimulus to market competition and technology neutrality. 

Cost-related motives have long been a main driver for OSS policy making, but less so in recent years~\cite{blind2021impact}. Motives for cost savings on the one hand relate to the ability to share and reuse existing OSS across PSOs, where one study has shown that organisations \textit{``would need to spend 3.5 times more on software than they currently do if OSS did not exist''}~\cite{hoffmann2024value}. On the other hand, there is potential for reduced license fees with proprietary software~\cite{deller2008open, cassell2008governments, koloniaris2018possibilities}, in addition to potentially free or low-cost maintenance for OSS~\cite{waring2005open}. Such ``free'' maintenance efforts, however, require an active and sustainable community backing the project, which cannot be taken for granted~\cite{linaaker2022characterize}, and support contracts may still be needed to enable a stable and secure adoption and use of the OSS alternative~\cite{shaikh2016negotiating}. 

In addition to cost savings, broader economic motivations include the potential for market stimulus and economic growth, as noted in several studies~\cite{ghosh2002free, ghosh2006study, greenstein2014digital}. Among more recent works, a study of a French national policy favouring OSS in procurement highlights how it led to a significant increase in OSS contributions and contributors per year from France, and how the subsequent growth of IT startups and the workforce impacted the country's economic growth~\cite {nagle2023government}. A similar study has also been performed on the EU context, highlighting how the workforce from member countries invested a corresponding amount of €1 billion in OSS, resulting in an impact of between €65 and €95 billion on the European economy~\cite{blind2021impact}. Korkmaz et al.~\cite{korkmaz2024github} also report similar findings in the US context.

The studies further highlight how increased investment had an impact on both GDP and entrepreneurial growth within the concerned countries. The latter has been further elaborated in follow-up studies showing that the number of new ventures increases with the number of contributions~\cite{wright2023open}.

\subsubsection{Political concerns}
Political concerns highlight both autonomy and independence, as well as transparency and democracy. Autonomy and independence have grown in importance in recent years and now form a central narrative in OSS policy globally~\cite{blind2021impact, lostri2023gov, jokonya2015investigating}.

Policies addressing these concerns commonly focus on managing dependencies and relationships with single vendors or technologies~\cite{deller2008open, lostri2023gov, blind2021impact}. When the source code is available as OSS, PSOs can define a preferred OSS solution in a tender and focus on selecting service providers rather than product suppliers. From a supply-side perspective, OSS can serve as a tool for opening closed markets by enabling new entrants to compete with incumbents~\cite{valimaki2005impact}.

Independence may also refer to the ability of PSOs to build their own internal capacity to develop and tailor services, tools, and infrastructure to their own needs and priorities, without being dependent on other PSOs or private actors~\cite{cassell2008governments, allen2010open}. Vendor independence and the ability to gain control, e.g., over upgrades, were the main drivers of the LiMux project, the City of Munich’s adoption of Linux-based operating systems~\cite{silic2017open}.

Autonomy and independence can also be a significant driver from an ideological perspective as illustrated in the case of Venezuela~\cite{maldonado2010process}. Here, OSS policy was part of a plan \textit{``for endogenous development intended to break Venezuelan dependence on foreign software and hardware. The main justification claims that previous governments spent more on licensing fees for proprietary software than on developing domestic technology and strengthening sovereignty, which have become top priorities for the Venezuelan socialist government''}~\cite{maldonado2010process}. 

A similar rationale can be observed, e.g., in Europe and the European Union in their strive towards achieving open strategic autonomy, able to make technical and sourcing decisions based on European values, norms, and laws~\cite{openforumeurope2022strategic}. Favouring OSS vis-à-vis proprietary solutions is considered a key step towards avoiding vendor lock-in and, by extension, strengthening digital sovereignty~\cite{blind2021impact}. The referred independence and sovereignty commonly also include promoting and growing a domestic or local ecosystem of vendors and service providers, investing in national R\&D capabilities, upskilling the digital workforce, and, by extension, boosting the economic growth of the corresponding region~\cite{bouras2014policy, blind2021impact, maldonado2010process}. Blind et al.~\cite{maldonado2010process} specifically note that Asian countries surveyed signal a greater focus on promoting OSS in their domestic industries, while European and American countries surveyed show a greater focus on public procurement of OSS-based solutions. 

Considering transparency and democracy, OSS may be seen as a means to scrutinise and promote trust in public digital services, what data they collect, and how this is processed~\cite{blind2021impact}. Citizens and society are, by extension, enabled to hold PSOs and policy-makers accountable, and demand change when needed~\cite{thevenet2024progress}. 

\subsubsection{Technical concerns}
Technical concerns are commonly expressed as Quality (or non-functional) requirements~\cite{blind2021impact}. Interoperability, standardisation, flexibility, customisability, reliability and scalability enabled by the openness of the source code are commonly cited~\cite{deller2008open, waring2005open}. These factors potentially enable increased efficiency in PSO processes, as systems may be tailored more closely to internal requirements~\cite{cassell2008governments, allen2010open}, thereby enabling further growth of e-government solutions~\cite{lakka2012does}. 

Together with open standards, OSS can further enable interoperable systems and public services within and across organisational borders~\cite{almeida2011open}. Being able to influence the pace of development is another factor highlighted~\cite{koloniaris2018possibilities, allen2010open}. Times of crisis, exemplified by the COVID-19 pandemic and the Russian invasion of Ukraine, demonstrate how OSS can enable the swift development and deployment of digital solutions by actors across sectors and borders~\cite{thevenet2024progress}. Blind et al.~\cite{blind2021impact} highlight improved access and maintenance of data stored in OSS using open standards as a main driver in recent years. 

\subsection{Policy implementations}
Issuing and communicating a policy is seldom sufficient to ensure its implementation and adoption by its target audience. In the case of OSS, the literature highlights several challenges, but also provides examples of how policy implementation can be supported. Below, we elaborate on both topics.
\subsubsection{Challenges and barriers}
Adoption and collaboration of OSS in the public sector does not come without friction. Several challenges are reiterated in the literature impacting the implementation of OSS policies and the realisation of their intended effects.

Culture is one area commonly highlighted, including a comfort with the status quo and resistance to change~\cite{persson2024soft}, as well as alternatives to established proprietary solutions~\cite{deller2008open, koloniaris2018possibilities}. Switching to an OSS-based solution may require multiple tools to be integrated, and not all functionalities may be present ``out of the box'', creating further resistance among end-users~\cite{cassell2008governments, koloniaris2018possibilities}. Risk-aversiveness may also be a factor in IT departments, even though OSS and proprietary solutions may co-exist~\cite{rossi2006study}. Koloniaris et al.~\cite{koloniaris2018possibilities} report from a Greek municipal context how the IT departments may be \textit{``sceptical and doubtful about the possibility of a radical change in the way they work and support their departments and the possibility that they would be requested to learn and support something totally new''}. Buy-in is required from all parts of the organisation, including the political, top-level, organisational, end-user, and IT~\cite{silic2017open}.

Securing leadership support, whether from local policymakers or management, is especially highlighted as cumbersome. Short-term horizons and risk aversion are commonly impeding characteristics. One reason is that adopting OSS may entail technical risks in tailoring and integrating identified components~\cite{cassell2008governments, silic2017open}. Existing relationships with established vendors, and lobbying from the same, further complicate matters~\cite{oram2011promoting}, including the blocking of potential policies in support of OSS adoption~\cite{chan2017coding}. Organisational aspects related to hierarchy and communication between leadership and other parts of PSOs are also key factors for success~\cite{cassell2008governments}. 

Lack of internal competence and resources needed to identify, evaluate and adopt the OSS is also cited as a common challenge~\cite{cassell2008governments, deller2008open}. This relates both to the technical know-how needed in software and requirements engineering~\cite{borg2018digitalization}, and to knowledge of how to adapt and apply the rigid procurement frameworks to which PSOs are bound~\cite {lundell2021enabling}. Misapplication of these frameworks can result in lock-in and costly migrations despite potential motives driving adoption, as well as costly overruling by competing vendors. Historical explanations can be found in how PSOs suffer due to a long tradition of outsourcing and the hiring of short-term consultants, which inhibit the growth of internal capabilities~\cite{marco2020outsourcing, cinar2019systematic}. Issues are especially apparent among PSOs limited in size and on the local levels of government~\cite{persson2024soft}.

Ensuring a sustainable availability of maintenance and support of the OSS project is also highlighted as a critical challenge~\cite{deller2008open}. Here, the literature reports that PSOs are sceptical about the level of trust that can be placed in the communities maintaining OSS projects~\cite{deller2008open, koloniaris2018possibilities}. Rather, they seem to express a need for vendors and service suppliers who can support the PSOs in adopting any OSS~\cite{lungo2007experiences}, aligning with the earlier-reported resource constraints. Still, there is scepticism about the availability of professional support at large, which may further inhibit consideration of OSS in the first place~\cite{cassell2008governments, koloniaris2018possibilities}.

\subsubsection{Support actions}

Aligning with the referred challenges, growing internal technical expertise within PSOs~\cite{oram2011promoting}, and creating champions who can support OSS adoption and collaboration, and integrate with OSS communities and wider ecosystems, are highlighted as success factors~\cite{van2015adopting}. The practice of having experts consult with and support entities within government on OSS matters has also been observed at the local level of the UK government and has been proposed as a policy recommendation in the study~\cite{shaikh2016negotiating}. 

The calls for internal experts and champions resonate with the increasing establishment of Open Source Program Offices (OSPOs) in later years, inspired by industry practice~\cite{haddad2020ospos}. The OSPOs are essentially support functions driving organisational change, supporting their overarching organisation in adopting and leveraging OSS in alignment with its business or policy goals~\cite{ruff2022rise}. They have been proposed in numerous policy recommendations~\cite{blind2021impact, nagle2022strengthening, herpig2023ospos, cisa2023ospo} and can help PSOs develop and apply the knowledge needed to interpret and implement relevant OSS policies. Linåker et al.~\cite{linaaker2023public} investigate and propose six archetypes for public-sector OSPOs based on a survey of PSOs in EU countries.

For PSOs with limited size and resources, however, creating their own OSPO may not be feasible. That is why collaboration between PSOs may be a way forward, both within~\cite{frey2023we} and across~\cite{marco2020outsourcing} different levels of government. Persson and Linåker~\cite{persson2024soft} describe how PSOs should work to establish joint OSS stewards, a form of proxy organisation that facilitates communication, collaboration, procurement, and maintenance of common OSS projects. These stewards would, in practice, fill the same role as an internal OSPO and support the owners or members of the steward. 

Examples of these kinds of stewards can be found, e.g., Denmark and Belgium with the organisations OS2 and IMIO, respectively. OS2, a Danish municipal association of 80+ out of 98 municipalities as members, demonstrates how PSOs can collaborate on the development and maintenance of common OSS projects, typically through procured services from their ecosystem of 60+ vendors and service providers~\cite{frey2023we}. Governance models, as well as procurement and development models, are standardised and signed off by involved vendors and service suppliers through Memorandums of Understanding. IMIO, on the other hand, is a service provider co-owned by 140+ Wallonian municipalities that develops and maintains common OSS projects based on requirements from its owners~\cite{viseur2023communesplone}.

As exemplified by the OS2 case, OSS stewards can further help ensure the availability of professional support. Shaikh~\cite{shaikh2016negotiating} highlights that it is essential \textit{``that a healthy ecosystem of small and medium sized firms is stimulated to service public sector open source products''} caring for the \textit{``detailed documentation, quality assurance, certification schemes, training, and support services''} also typically required for proprietary products and services, although already included in product licenses and service contracts. The OSS stewards and OSPOs in general can further ensure that vendors and service providers \textit{``agree [to] the open source approaches, so that source code can be reused across the government''} and perform any development tasks in an open, agile and collaborative manner as expressed by Mergel~\cite{mergel2016agile}. 

\subsection{OSS indicators in Digital Government Indexes}
Since the 1970s, international rankings and benchmarks \textit{``by which something can be measured or
judged''}\footnote{\url{https://www.merriam-webster.com/dictionary/benchmark}} have become pervasive~\cite{merry2011measuring}. Research on how governments respond to international rankings suggests that these can cause political and economic change as much as reflect it. Governments often view high rankings as a signal of legitimacy, which can lead to targeted reforms~\cite{broome2018bad}. Rankings can also heighten status concerns, particularly when compared with peer or rival nations, encouraging compliance to avoid reputational risk and enhance global standing~\cite{kelley2015politics}. While there is a risk that some states, especially those less economically independent, adopt specific policies to improve their ranking metrics without big structural changes, these rankings can nevertheless be an important driver of policy change.

Indicators are commonly adopted in the context of Open Data~\cite{zuiderwijk2021comparing}, but much less so in the context of OSS. One example is the Principles for Digital Development~\cite{dial2024principles}, developed with input from USAID and other international bodies, which emphasise the use of open standards and collaborative design as best practices across digital project stages. The OECD’s Digital Government Policy Framework echoes this by advocating OSS adoption to avoid vendor lock-in, enhance service interoperability, and foster a competitive digital public sector~\cite{oecd2024framework}. The OECD’s Digital Government Index asks for the \textit{``availability of guidelines to use open source to develop digital government initiatives''}, and \textit{``available actions related to the use of open source''}~\cite{oecd2023index}.

The Global Innovation Index (GII), published by the World Intellectual Property Organisation (WIPO), now includes metrics that reflect contributions from OSS developers, such as GitHub commits~\cite{wipo2024index}. This update further recognises OSS development as a significant form of innovation alongside traditional indicators such as patents and academic publications.

Other recognized digital government indexes show no to limited attention to OSS, including the Digital Economy and Society Index (DESI), tracks EU countries’ digital performance across themes such as connectivity, digital skills, the digital transformation of businesses, and public services~\cite{ec2024desi}, and the eGovernment Benchmark\footnote{\url{https://digital-strategy.ec.europa.eu/en/library/digital-decade-2024-egovernment-benchmark}}, used to measure user centricity, transparency, key enablers, and cross-border services across Europe~\cite{ec2024egov}. The UN E-Government Survey mentions OSS in the context of Digital Public Goods as a means of achieving the global Sustainable Development Goals (SDGs) and promoting interoperability, but provides no explicit indicator~\cite {un2024egov}. The UNDP Development Compass considers collaboration and sharing in broader terms under the umbrella of open government\footnote{\url{https://digitaldevelopmentcompass.undp.org/methodology/digital-development-compass}}.

Although some indicators are available, they are limited and provide little guidance to governments and public sector entities on how to consider and leverage OSS in the planning, implementation, and follow-up of their digital transformation programs. As many governments look to digital maturity indexes, neglecting markers related to the adoption of OSS as a policy instrument may have serious negative implications and result in lost opportunities.

\subsection{Summary}
OSS policies have been widely adopted globally, focusing on research, development, and procurement~\cite{lostri2023gov}. These policies vary in formality and scope, ranging from parliamentary bills to informal expressions of support~\cite {blind2021impact, thevenet2024progress}. Countries like Venezuela have mandated OSS adoption~\cite{maldonado2010process}, while others, such as France, have successfully integrated OSS into procurement processes (Nagle, 2023). Motivations for these policies include economic benefits like cost savings~\cite{deller2008open, cassell2008governments} and market growth~\cite{blind2021impact, nagle2023government}, political goals such as autonomy and transparency~\cite{deller2008open, lostri2023gov}, and technical advantages like interoperability and flexibility~\cite{thevenet2024progress, bouras2014policy}. 

Challenges in OSS adoption include cultural resistance~\cite{persson2024soft}, leadership support~\cite{van2015adopting}, competence gaps~\cite{cassell2008governments}, and ensuring sustainable maintenance~\cite{deller2008open}. Strategies to overcome these challenges include establishing Open Source Program Offices (OSPOs)~\cite{linaaker2024public} and collaborative stewardship models like Denmark's OS2~\cite{frey2023we}, which aim to enhance internal expertise, foster collaboration, and ensure professional support for OSS projects. 

Despite the widespread reporting of practice, there is a limited systematic overview of policy design and support initiatives. The availability of OSS enablement indicators in established digital government indexes that provide guidance and benchmarking is also limited to non-existent. These are the gaps we aim to address through the present study.

\section{Research design}
\label{sec:researchdesign}
We adopted a multiple-case study design to explore the phenomena in their real-world context, using the case study methodology proposed by Runeson et al.~\cite{runeson2012case}, in turn based on the work by Yin~\cite{yin2009case}. While these guidelines were developed more than a decade ago, they are still broadly adopted and relevant for the software engineering research today\footnote{\url{ https://www2.sigsoft.org/EmpiricalStandards/docs/standards?standard=CaseStudy#}}. The methodology offers the possibility to gather in-depth knowledge and develop an understanding of its occurrence, and is particularly suitable when contextual factors cannot be removed or isolated. We adopt a pragmatic lens in the research, leveraging multiple sources of knowledge to generate new insights and qualitatively synthesising a common understanding across the investigated cases.

\subsection{Case sampling}
We adopted a purposive sampling approach~\cite{patton2014qualitative}, with the goal to identify mature countries in terms of digital maturity. The aim was to investigate their policy and related support actions related to the use of OSS as an instrument for software reuse and collaborative development, which we consider our units of analysis. To support our sampling, we identified four international indices for digital maturity.

\begin{itemize}
    \item The Digital Economy and Society Index (DESI)~\cite{ec2024desi}
    \item eGovernment Benchmark~\cite{ec2024egov}
    \item UN E-Government Survey~\cite{un2024egov}
    \item OECD Digital Government Index\cite{oecd2023index}
\end{itemize}

Taking maturity, geographical representation, and resource constraints into account, 16 countries were sampled, as presented in Table~\ref{tab:countries-sampled}. Fourteen of these were selected because they were among the top ten in at least two listings. Four were chosen because they were among the top five in at least one list. One additional country was sampled to improve geographical representation in the South Pacific (New Zealand), and one additional country was sampled to ensure representation of an earlier reported mature OSS adopter (France).

\begin{table}[]
\caption{Overview of countries sampled, and main rationale.}
\label{tab:countries-sampled}
\begin{tabular}{p{2cm}p{5.5cm}}
\toprule
\textbf{Country} & \textbf{Rationale} \\ \midrule
Denmark & Top 10 in at least two lists. \\
Estonia & Top 10 in at least two lists. \\
Finland & Top 10 in at least two lists. \\
Iceland & Top 10 in at least two lists. \\
Korea & Top 10 in at least two lists. \\
Luxembourg & Top 10 in at least two lists. \\
Malta & Top 10 in at least two lists. \\
Spain & Top 10 in at least two lists. \\
Sweden & Top 10 in at least two lists. \\
The Netherlands & Top 10 in at least two lists. \\
United Kingdom & Top 5 at least in one list \\
Colombia & Top 5 at least in one list \\
Japan & Top 5 at least in one list \\
Ireland & Top 5 at least in one list \\
France & Top 10 in at least one list, mature user of OSS \\
New Zealand & Top 10 in at least one list, geographical representation of the South Pacific \\ \bottomrule
\end{tabular}
\end{table}

\subsection{Theoretical framework}
A theoretical framework was developed covering the two dimensions of policy and policy support actions. The framework draws on the comprehensive report by Blind et al. for the European Commission, which investigates the impact of OSS on the European economy~\cite{blind2021impact}. The report conducted a comprehensive analysis of several countries regarding how OSS policies are formalised, based on the policy literature~\cite{blind2021impact}. The design choice was purposive, aiming to strengthen the validity of this study and allow connection of the results between the two works (while recognising differences in contextual factors, etc.). Our work extends that of Blind et al., most notably by adding a dimension on support actions.

\begin{table*}[!htbp]
\caption{Theoretical framework used in the study, developed covering the two dimensions of policy, and policy support actions. The framework draws from earlier reports investigating best practices of OSS in the public sector~\cite{blind2021impact, linaaker2024public}.}
\label{tab:questions}
\begin{tabular}{p{4cm} p{12.5cm}}
\toprule
\textbf{Area} & \textbf{Question} \\  \midrule
\multicolumn{2}{c}{\textbf{Dimension: Policies}} \\ \midrule
\textbf{Type of policy} & 
    * Are there any national policies or strategies prescribing the (re)use, sharing and collaboration of software and OSS specifically? Consider:
    
     - General strategies for e-government services and internal use
     
     - Specific domain, e.g., Science, Employment Service, Digital Infrastructure, Public procurement.
     
      - Digital sovereignty, i.e., avoidance of lock-in to any specific format, platform, technology, or vendor, and being able to make technological decisions based on national laws, values, and needs?
      
      - Cybersecurity aspects related to OSS used with the public sector or society at large?
\\
\textbf{Scope and purpose of policy} & 
    * What is the scope and purpose of these policies or strategies?
    
    * Is it internally on the focal administration, and/or externally focused on directing and supporting external organizations?
    
    * Where is it executed and enforced? E.g., level of government?

\\
\textbf{Stakeholders} & 
    * In which ministries or PSOs are the strategies anchored? 
    
    * How is it enforced and realized? 
    
    * Which stakeholders are involved?

\\
\textbf{Prescriptions} & 
    * What policies, recommendations, or guidelines are given related to software (re)use, sharing and collaboration of software and OSS specifically?
    
    * Are the policies, recommendations, or guidelines advisory (recommended), preference (preferred but not mandatory), or mandatory (required)?

    * Consider both acquisition of solutions (no procurement), procurement of products or services, and internal or collaborative development perspectives.

\\ \midrule
\multicolumn{2}{c}{\textbf{Dimension: Policy support actions}} \\ \midrule
\textbf{Current state and use of OSS} & 
    * What role does software reuse through OSS play in the country?
\\
\textbf{Organizational Support} & 
    * Are there any formal or informal centers of competency (similar to an Open Source Program Office (OSPO)) supporting the adoption of OSS and reuse of software?
    
    * On what mandate and policy are they acting? 
    
Compare and align with policies listed earlier.

    * What is their scope and purpose?
    
    * Is it internally on the focal administration, and/or externally focused on directing and supporting external organizations?
    
    * Where is it executed and enforced? E.g., level of government?
    
    * How are these they organized and structured? 

    * Consider OSPO archetypes described in the EC OSPO study: National Government, Local Government, Association-based, Institution-centric, Academic, and Independent OSPOs.

\\
\textbf{Funding} & 
    * Is there any funding or state aid provided for promoting or enabling the (re)use, sharing and collaboration of software and OSS specifically?
    
    * Are there any additional types of support provided?

\\
\textbf{Development and release} & 
    * How is development, governance, and ownership of intellectual property related to software released as OSS managed?
    
    * Are there any policies, recommendations, guidelines, or best practice in place?
    
    * How is long-term maintenance, quality, and sustainability of OSS considered and ensured?

\\
\textbf{Software inventory} & 
    * How are solutions inventoried and promoted for reuse, and collaboration?

    * Consider any public software catalogs or use of external social coding platforms such as GitHub and GitLab.

\\
\textbf{Promotion} & 
    * Are there any organized or informal activities promoting or enabling the (re)use, sharing, and collaboration on software and OSS specifically?
\\
\textbf{Reuse} & 
    * How is the (re)use, sharing, and collaboration of OSS perceived across public sector organizations, and levels of government?
    
    * What actions are being made to improve the (re)use, sharing, and collaboration of software, and OSS specifically?

\\ \bottomrule
\end{tabular}
\end{table*}

\subsection{Data collection and analysis}
Data for each country were initially collected through desk research, including policy and regulatory documents, public reports and investigations, and grey and academic literature. The country intelligence reports published by the Open Source Observatory~\cite{osor2025country} served as a starting point for many European countries. The policy analysis of individual countries presented in Blind et al.~\cite{blind2021impact} also provided essential input and starting points for the research.

Each author conducted one case study in parallel, for a total of eight per author. For each case, the investigation alternated between the two policy dimensions and policy support actions addressing questions defined by the theoretical framework in Table~\ref{tab:questions}, as new documentation was identified. Investigations were typically initiated by identifying legislative and regulatory frameworks through the aforementioned starting points. These were structured by their chronological introduction into their respective national contexts, along with any guidelines, recommendations, and public documentation reporting on policy actions implemented. Online searches on Google and the Joinup Open Source Observatory news archive were used to complement our country-specific investigations.

Texts were extracted and summarised for each policy entity and policy support action. After the desk research was considered to have reached saturation, texts were synthesised into a narrative structure, suitable for a practitioner as a country report. The country report was sent to two current or former government officials with in-depth knowledge about the state of OSS policy and related support actions in the concerned country. Individuals were sampled via organisations (typically national-level public authorities responsible for digital government at large or for digital transformation in a specific field, such as public employment) identified through the desk research conducted for the country reports.

This was followed up with semi-structured interviews, guided by the theoretical framework defined in Table~\ref{tab:questions}, and related to the communicated country report to discuss its validity and to add, change, or retract any details as needed. Each interview lasted about 60 minutes and was performed via online video conferencing platforms. A subset of interviews was not recorded due to interviewees' privacy concerns. When available, recordings were used to verify notes and interpretations. Note-taking was performed during the interview, and summarised afterwards, and cross-checked with interview recordings when available. The country reports were revised accordingly and communicated to the interviewees to validate the interview takeaways and finalise the country reports. 

Case data were analysed using a narrative synthesis approach as proposed by Cruzes et al.~\cite{cruzes2015case}, in turn based on earlier works in the health domain~\cite{Popay2006NarrativeSynthesisGuidance, popay2010narrative, pope2007synthesizing}. The approach is structured into four steps: 1) theorisation and conceptualisation, 2) preliminary analysis per case, 3) exploration of relationships within and across cases, and 4) robustness assessment of the synthesis~\cite{cruzes2015case}. The theorisation and conceptualisation step generated the theoretical framework as presented in 3.2. The following preliminary case-based analysis was conducted iteratively, with case data structurally coded using the areas defined in Table~\ref{tab:questions} as an a priori codebook. High-level codes were iteratively identified, merged and synthesised into higher-level themes using open and axial coding, e.g., the different dimensions of policy documents (inbound vs outbound direction, and internal vs external focus). The posterior codebook is presented in Tables~\ref{tab:policy} and~\ref{tab:support-actions}, respectively. The codes were iteratively discussed among the authors, taking into account their respective cases, to reach agreement on the codes and their definitions. Country reports are available in the report underpinning this study~\cite{linaaker2024reuse}. 

After the case-based analysis, relationships within and across the 16 cases were explored based on the high-level themes (Tables~\ref{tab:policy} and~\ref{tab:support-actions}), which are presented in the narrative synthesis in Section 4. Finally, a set of indicators was derived from the higher-level themes (Tables~\ref{tab:policy} and~\ref{tab:support-actions}) that may provide guidance on planning, implementation, and follow-up for the development of policy and related support actions for OSS as an instrument for software reuse and collaborative development. The indicators are formulated to be measurable and compared annually. Formulation and definition are inspired by language and structure from established digital government indexes~\cite{ec2024desi, ec2024egov, un2024egov, oecd2023index}. 

The last step in the narrative synthesis process, robustness assessment of the synthesis~\cite{cruzes2015case}, was conducted through an additional round of member-checking, during which interviewees were provided with the full report, resulting in only minor feedback on the synthesis. Regarding the indicators, a recurring comment was that the ones proposed are too extensive to implement in their entirety and should instead be seen as a source for eliciting indicators tailored to need and context (as highlighted in the relevant section). Comments further noted that some indicators may be difficult to compare across countries due to differences in country size and government (the referenced indicators have been revised accordingly). Some indicators were completely removed, including a section on "Public sector networks for OSS knowledge sharing," based on feedback highlighting their benefit but limited value as indicators. A topic for future work is to further prioritise the suggested indicators and to prototype and evaluate questionnaires that enable real-world adoption and implementation. 

\subsection{Limitations and threats to validity}
We adopt the criteria for naturalistic inquiries proposed by Guba~\cite{guba1981criteria} for discussing the limitations and threats to validity of our work. These criteria include credibility, transferability, dependability, and confirmability.

\textbf{Credibility} concerns the extent to which the findings accurately represent the realities studied. To strengthen credibility, we triangulated data from multiple sources, combining desk research of policy documents with semi-structured interviews across all cases. The iterative development of country reports, followed by member-checking with knowledgeable government representatives, further helped validate interpretations and reduce misrepresentation. However, the number of interviewees per country was limited, typically to two participants due to resource constraints, which may restrict the breadth of perspectives captured. The case studies were, hence, focused on breadth rather than depth, aiming to get a broad overview of how policy and support actions can be formulated and organised. 

\textbf{Transferability} refers to the extent to which the findings can be applied in other contexts. This study focuses on 16 digitally mature countries selected through purposive sampling based on international digital government indexes. While this enables in-depth exploration of advanced practices, it also implies that the findings may not be directly transferable to less mature contexts or to countries with substantially different institutional, cultural, or political conditions. We mitigate this limitation by providing rich descriptions of the cases and their contexts, enabling readers to assess the applicability of findings to their own settings. The proposed indicators should therefore be interpreted as context-sensitive suggestions rather than universally applicable prescriptions.

\textbf{Dependability} relates to the consistency and repeatability of the research process. We employed a structured case study methodology and a clearly defined theoretical framework to guide data collection and analysis. The use of a shared codebook, iterative coding, and cross-author discussions contributed to consistency in interpretation across cases. Nevertheless, qualitative analysis inherently involves researcher judgment, and the iterative refinement of codes and themes may introduce variability that could affect replicability. Furthermore, the evolving nature of policy landscapes means that findings represent a snapshot in time, and subsequent developments may alter the observed dynamics.

\textbf{Confirmability} addresses the degree to which the findings are shaped by the respondents and data rather than researcher bias. To enhance confirmability, we maintained transparent links between data sources, interpretations, and synthesised findings, and we provided references to original materials where possible. Member-checking of both case reports and synthesised results further grounded interpretations in practitioner perspectives. However, the study scope excluded actors outside the public sector, such as vendors, civil society, and broader OSS ecosystems, which may play a significant role in shaping policy outcomes. This delimitation, along with reliance on publicly available documents and selected interviewees, may affect the completeness of the analysis.

\section{Findings}
\label{sec:findings}

\subsection{Policy and stakeholders}
\begin{sidewaystable*}[!htbp]
\caption{Overview of policy dimensions and types, along with definitions and references to examples derived through thematic synthesis of the case studies.}
\label{tab:policy}
\begin{tabular}{p{3cm}p{3cm}p{13cm}p{3cm}}
\toprule
\textbf{Policy dimension} & \textbf{Policy type} & \textbf{Description} & \textbf{Examples} \\ \midrule
\multirow{2}{*}{\textbf{Policy focus}} & Internal & Focus on the PSO’s own use of or contribution to OSS. & Estonia, France, the Netherlands \\
 & External & Aimed at encouraging OSS uptake in the private sector. & Japan, South Korea, Colombia \\
\multirow{2}{*}{\textbf{Policy direction}} & Inbound & Concerns acquisition and procurement of OSS for internal purposes. & Malta, France, Spain \\
 & Outbound & Pertains to the release of software developed through public funds. & Iceland, New Zealand, Denmark \\
\multirow{3}{*}{\textbf{Type of intervention}} & High-level endorsement & High-level endorsements within policy documents of a more general nature encouraging the use or release of OSS. & Colombia, Finland, Luxembourg \\
 & Advisory & Recommend considering, comparing, and evaluating OSS on an equal footing with proprietary alternatives in acquisition and procurement policies (inbound context) and as a mechanism for releasing and reusing software developed with public funds (outbound context). & Denmark, Iceland, Malta \\
 & Prescriptive & Explicitly expresses a preference for OSS before other alternatives unless special circumstances apply in the acquisition and procurement process (inbound context), or a preference is for releasing public sector software as OSS by default, unless specific considerations, such as security or confidentiality, dictate otherwise (outbound context). & France, the Netherlands, Spain \\
\multirow{3}{*}{\textbf{Form for definition}} & Legislative & Defined in rule of law adopted by the national legislative body. & France, the Netherlands \\
 & Government instruction & Defined in formal instructions from the national government. & France, UK \\
 & Strategy documents & Defined in national or agency-specific strategies, e.g., related to digital government, transformation, or procurement. & Iceland, New Zealand \\
\multirow{3}{*}{\textbf{Scope of policy}} & National government & Regards PSOs on the national level of government. & France, Spain, Iceland \\
 & Regional and local government & Regards PSOs, regional and/or local levels of government. & Denmark, Spain \\
 & Institution-specific & Concerns specific PSOs. & Sweden \\
\end{tabular}
\end{sidewaystable*}

As approached in this study, government policies refer to a set of principles, objectives, and guidelines explicitly formulated by a government or other authoritative bodies at the national level and designed to influence or determine decisions and actions, or to offer guidance. The policies analysed in this report are aimed at improving the conditions for the use and reuse of software, with OSS viewed as a mechanism for advancing such practices. While adopted policies do not always reflect the actual practice and impact of software reuse through OSS in the country, they serve as evidence of political-level awareness of the value of such practice in advancing desired outcomes in the public interest.

Across a diverse sample of 16 countries, we observed notable variations in policies concerning scope, objectives, and levels of prescription. Despite this diversity, which at least in part can be attributed to different institutional frameworks, distinct groupings emerged based on several discernible criteria. In the following section, we compare and contrast the policies according to these criteria to provide insights into both commonalities and divergences. The resulting categorisation, summarised in Table~\ref{tab:policy}, provides a more nuanced understanding and a basis for drawing conclusions and offering recommendations in subsequent sections.

\subsubsection{Focus of policies}
A first distinction can be made between government policies that focus on the PSO’s own use of or contribution to OSS (internal focus) and those that aim to encourage OSS uptake in the private sector (external focus). The overall emphasis in this study is on the former category, and most of the countries in the sample have policies of that nature. Yet the two Asian countries included in the analysis, South Korea and Japan, are notable exceptions, as their governments have adopted several policy measures since the early 2000s to actively encourage and support OSS uptake in their domestic tech industries~\cite {koreangov2020software, japanministry2022collection}. Such activities can also be observed to a lesser extent in Colombia, where a national program promotes OSS use by SMEs~\cite{worldeconomicforum2021data}.

The internally focused policies can be further divided into those focused on the use (or consumption) of OSS and those centred on its development and release processes. The former concerns the acquisition and use of OSS, while the latter concerns the release of software developed with public funds. Policies addressing the use and adoption of OSS are here referred to as inbound OSS policies, while those concerning the development and release of OSS are termed outbound OSS policies. The inbound and outbound contexts may be addressed by distinct policies or by joint policies that address both use cases.

In France~\cite{secretariat2012Orientations, loi2016}, and the Netherlands~\cite{dutchgoc2023wet, dutchgov2022inboundpolicy}, these aspects are considered in separate policies, whereas in Malta~\cite{maltagov2019inboundoutbound} and Iceland~\cite{icelandgov2021inboundoutbound} they are addressed jointly. It should be noted that the borders between these policy domains are not always clear-cut, and the distinction appears to be linked to the evolution of policies and the maturity of OSS use. Policies adopted in the early 2000s, such as in the UK~\cite{ukgov2002inboundoutbound}, focused on inbound consumption (procurement), whereas more recent policies, such as in Estonia~\cite{estoniangov2021inboundoutbound}, have shifted to include outbound consumption. In some countries, such as Colombia~\cite{columbiagov2019inbound}, the focus is almost exclusively on promoting increased use of OSS, with no outbound direction or guidance.

Where these policies are considered separately, they may also be owned by different parts of the government, and the intervention may be based on distinct arguments or justifications. For instance, inbound policies may reside in procurement rules under a ministry of finance to ensure responsible use of government funds, while separate outbound policies may be housed in a department responsible for digital transformation, driven by principles of open innovation (more examples of how these factors interact with each other are provided below).

\subsubsection{Type of intervention}
The policies examined in this report also vary in terms of the type of policy measure and the degree to which software reuse through OSS is prescribed. In a first group of countries, government intervention regarding OSS mainly takes the form of high-level endorsements in policy documents of a more general nature. Examples include Colombia, which in its 2018-2022 and 2022-2026 National Development Plans~\cite{columbiagov2018nationaldevplan, columbiagov2022nationaldevplan} mentions promoting OSS. Similarly, recent government programs in Finland~\cite{finnishgov2019govactionplan} and Luxembourg~\cite{luxgov2018policy} have outlined the administration's intention to encourage OSS uptake, but these commitments have not, to a significant degree, been translated into concrete guidance documents or specific national-level policies.

A second distinct group comprises countries that have adopted explicit OSS advisory policies. These policies recommend considering, comparing, and evaluating OSS on an equal footing with proprietary alternatives in acquisition and procurement policies (inbound context) and as a mechanism for releasing and reusing software developed with public funds (outbound context). Notable examples include Denmark~\cite{danishgov2022vejledning} and Iceland~\cite{icelandgov2021inboundoutbound}, where advisory policies encourage the adoption of OSS for both (re)use and release of public software.

Lastly, a distinct group of policies explicitly expresses a preference for OSS. In the inbound context, these policies require that OSS be selected before other alternatives unless special circumstances apply during the acquisition and procurement process. In the outbound context, the preference is for releasing public sector software as OSS by default, unless specific considerations, such as security or confidentiality, dictate otherwise.

In the UK, while rules have evolved separately, some policy documents integrate both aspects, creating an expectation that OSS is the default option without specific mandatory legislation for inbound or outbound~\cite{ukgov2022inboundoutbound, ukgov2023inboundoutbound}. France~\cite{loi2016} and the Netherlands~\cite{dutchgoc2023wet} have outbound policies mandating that public sector software be released as OSS unless special circumstances apply, e.g., in relation to security, confidentiality, or integrity aspects, while both have advisory policies for the inbound context. In Spain, all PSOs are required to release any public sector software for internal reuse within the government, and, if deemed appropriate, use OSS to enable such reuse~\cite{spanishgov2015outbound}. All PSOs are correspondingly obliged to consider any public software, OSS or not, in the initiation of any acquisition and procurement process. Estonia has no general inbound policy but has recently adopted a law stipulating that all software developed with taxpayers’ money should be published with an OSS license unless doing so would harm national security~\cite{estoniangov2021inboundoutbound}. 

\subsubsection{Scope}
The policies differ further in terms of whom they apply to or address. The French inbound policy, issued as a government instruction, addresses PSOs at the national level of government, while the outbound policy, as defined by the law, applies to all PSOs~\cite{secretariat2012Orientations}. In Denmark, national policies are defined in guiding architectural documents~\cite{danishgov2022vejledning}, on the one hand, by the Agency for Digital Government addressing all PSOs on the national level of government, while on the other hand, a corresponding policy is provided by the Association of Regions and Municipalities applying to PSOs on the regional and local levels of government~\cite{danishgov-KL2022vejledning}. In contrast, policies and guidelines found in Sweden are typically limited to single PSOs~\cite{digg2022policy, forsakringskassan2019riktlinje}. In the UK, the inbound policy "playbook" applies directly to central government agencies on a ‘comply or explain’ basis and is to be considered ‘good practice’ by the wider public sector~\cite{ukgov2022inboundoutbound, ukgov2023inboundoutbound}.

\begin{framed}\noindent\textbf{Summary}: Most surveyed countries have established policies addressing software reuse through OSS, encompassing both inbound (acquiring new software) and outbound (sharing acquired solutions). While policies differ in scope and level of prescriptiveness, they are, in most cases, owned by central PSOs with responsibility for areas such as digital government, transformation, and procurement. The main emphasis in this study is on policies concerning the public sector’s own use of OSS, yet in a notable subset of countries, these policies had an external focus aimed at increasing the uptake of OSS in the domestic technology sector.
\end{framed}

\subsection{Policy goals}

\subsubsection{Economic factors – OSS to avoid double spend, lock-in, and promote a competitive market}
Present in all initiatives is some notion of encouraging responsible public spending and reducing lock-in to specific vendors. The potential for cost savings and efficiencies was particularly prominent in earlier policies and is seldom provided as the only reason for promoting OSS in more recent policy documents.

Within the general focus on economics, distinct arguments are made. For example, the principle that the public sector should not pay for the same solution twice is explicit in policies that require individual contracting authorities to acquire rights to allow reuse within the public sector. Examples include Colombia~\cite{columbiagov2019inbound}, Spain~\cite{spanishgov2015outbound}, and the UK~\cite{ukgov2023inboundoutbound}.

The (re)use of OSS is also seen as a means to increase competition among suppliers in a procurement process. As the source code and, ideally, all necessary knowledge and infrastructure are openly available, suppliers unfamiliar with OSS can enter the market, although a knowledge barrier may still exist. Studies in France and across Europe both show the potential for increased competitiveness, growth in small- and medium-sized companies, and a positive impact on GDP~\cite{nagle2019government, blind2021impact}.

\subsubsection{Interoperability – OSS as a mechanism for interoperable infrastructure and public services}
The European Interoperability Framework (EIF) and National Interoperability Frameworks (NIFs) have also proved to be important impetuses for several of the policies. The motivation is often combined with other value drivers such as cost efficiencies and innovation, but technical interoperability is, in some cases, seen as an overarching driver for reuse and the adoption of OSS.

In Estonia, the decision to use OSS appears to have been driven by technological pragmatism and the need to accelerate its digital transformation, building on existing components while ensuring interoperability across government agencies. Recently, the Estonian government has recognised the value of tapping into a global community of developers~\cite{estoniangov2021inboundoutbound}.

In Spain, the National Interoperability Framework underpins legislation requiring PSOs to share and reuse public sector software as far as possible, where OSS is seen as a mechanism for reuse if such a release contributes to greater transparency in the PSO’s operations~\cite{spanishgov2015outbound}. Although not as explicit, other countries, such as New Zealand~\cite{newzealandgov2016inboundoutbound}, Sweden~\cite{esam2022råd}, and Iceland~\cite{icelandgov2021inboundoutbound}, also explicate the value in promoting interoperability and harmonisation across public services and the national digital infrastructure.

\subsubsection{Digital sovereignty – OSS as a means to empower sovereign decisions on the use of technology}
Digital (or technical) sovereignty highlights the importance and means of making technical sourcing and design decisions in line with local law, norms, and values. In France, digital sovereignty is implicitly highlighted as a policy goal through the Digital Republic law, which states that administrations shall ensure that their information systems remain under control, sustainable, and independent~\cite{loi2016}.

In Sweden, digital sovereignty is also implicitly mentioned in several PSO-specific policies~\cite{digg2022policy, forsakringskassan2019riktlinje}. The general discourse on the topic, however, has received significant attention in debates on cloud and data management. eSam, a national collaboration between 30+ PSOs, is, for example, driving an investigation into possible communication and collaboration tools allowing for hosting and data management in line with European legislation in the area~\cite{esam2022råd}. Private vendors have now begun packaging services based on different OSS solutions, such as Nextcloud for document management, Element for chat, and Jitsi for video conferencing. The Swedish Insurance Agency and the Tax Agency are also investigating a public-sector alternative for the corresponding solutions. Looking beyond the surveyed countries, Germany also provides a similar example through the development of their OpenDesk solution, a compilation of OSS-based solutions aiming to provide a sovereign option to the desk suite for civil servants, including the collaboration and communication tools surveyed and implemented in Sweden.

Communication is also an important area in Luxembourg, where digital sovereignty has been invoked as a rationale for specific initiatives, such as the development of LuxChat, an OSS instant messaging service for the public sector developed in partnership with an ecosystem of several providers to safeguard the proper use of data~\cite{luxgov2018policy}. In France, a corresponding alternative is developed through the Tchapp project.

The Basque Country, a region in Spain, provides an example of a transition to OSS-based tools and infrastructure that has matured to the point where all the public sector uses OSS-based operating systems and productivity suites~\cite{basquegov2018catalog}. A partial motive has also been to localise the software to the regional language, further increasing the sense of independence in the region.

Policies in Japan and Korea, the two Asian countries included in the sample, have been formulated with the clear aim of supporting technological independence~\cite{japanministry2022collection, koreangov2020software}. In contrast with the rest of the countries in the sample, OSS promotion is aimed at the private sector as part of an industrial strategy. Korea, in particular, has invested significant resources and built institutional competence, not to guide public sector users but to support uptake in its tech sector. 

\subsubsection{Security – OSS as a (potentially) robust building block in need of maintenance}
Security in OSS is commonly discussed in both positive and negative terms. One discourse emphasises the risks associated with having source code openly available, potentially exposing vulnerabilities to identification, introduction, and exploitation. Another perspective views OSS as robust and secure, leveraging transparency to enable multiple eyes to review the source code, thereby identifying and addressing issues early and reducing the risk of vulnerabilities. The security of OSS depends on its sustainability—how well-maintained it is over time, without disruptions or quality degradation.

While many policies emphasise the importance of a functioning, interoperable digital infrastructure free of vendor lock-in, there's often limited attention to the sustainability of the OSS building blocks that underpin it. France is an exception, where the government instruction Circulaire 5608 recommends dedicating 5-10 per cent of any funds saved through an OSS-related acquisition to contribute back to the concerned OSS projects and their dependencies~\cite{secretariat2012Orientations}.

The emphasis on sustainability is often found in guidelines that help implement and realise defined policies. In Sweden, many PSO-specific policies and guidelines emphasise the importance of contributing changes or additions back to OSS projects~\cite{digg2022policy, forsakringskassan2019riktlinje}. The Netherlands also emphasises this through government-commissioned reports~\cite {dutchgov2022inboundpolicy}. In France, the guidelines and support from the National government OSPO focus on encouraging contributions back to OSS projects and their further development~\cite {dinum2023doc}.

The level of security and trust in OSS is further highlighted through its adoption and use among cybersecurity agencies, such as the House of Cybersecurity in Luxembourg~\cite{luxhousecybersecurity2023open} and the National Agency for the Security of Information Systems (Agence nationale de la sécurité des systèmes d'information – ANSSI) in France~\cite{anssi2023opensource}. Both actively (re)use OSS and participate in the collaborative development of several tools. ANSSI also has an explicit, diverse approach to promoting and supporting the sustainability of several core OSS projects of both internal and national interest.

In Japan, the government has established a software security task force to address private-sector use of OSS~\cite{japanministry2022collection}. It has published guidelines for appropriate software management methods and responses to vulnerabilities and license issues.

\subsubsection{Transparency – OSS as an enabler for trust, control, and innovation}
Transparency is a recurring theme in many policies. In France and the Netherlands, transparency is a driving factor in their outbound policies, enacted in the legislation of their respective countries~\cite{loi2016, dutchgoc2023wet}. This effectively treats source code as public data and administrative documents that should be released upon public request. In the Netherlands, this approach is a response to earlier incidents in which algorithms used in public services produced discriminatory recommendations.

Similarly, in Colombia~\cite{columbiagov2019inbound}, Sweden~\cite{digg2022policy}, and New Zealand~\cite{newzealandgov2016inboundoutbound}, the use of open technologies is expressed as a way to enhance trust between the government and other stakeholders, including citizens. In Spain, the potential to increase transparency in government services is explicitly cited as a factor to consider when deciding whether a public-sector software project should be released as OSS~\cite{spanishgov2015outbound}.

In some cases, these policies are part of a broader push for open government and open innovation. Luxembourg, for example, views OSS as a means to enable the co-creation of government services by involving both public and private actors~\cite{luxgov2018policy}. This reflects a broader trend toward openness and collaboration in the development and provision of government services.

\begin{framed}\noindent\textbf{Summary}:
The rationale for introducing government policies that promote software reuse through OSS in the public sector stems from several factors. Policy documents typically draw on several such factors to make the case for encouraging OSS. Economic factors are a driver in almost all cases, aiming to avoid double-spending and vendor lock-in and to foster market competition. Digital sovereignty is highlighted in some countries and is driving specific initiatives. Security considerations emphasise the dual perspectives of risk and opportunity that transparency provides and, in some cases, highlight the need to support and contribute to the maintenance of critical OSS components used in digital infrastructure. The benefits of transparency are further discussed, e.g., in terms of data collection and management, algorithm-based decision-making, and defining interfaces for third-party actors to interact with.
\end{framed}

\subsection{Implementation and support}

\begin{table*}[!htbpt]
\caption{Types of support actions related to the implementation of OSS policies.}
\label{tab:support-actions}
\begin{tabular}{p{2cm}p{12cm}p{2cm}}
\toprule
\textbf{Support action} & \textbf{Description} & \textbf{Example} \\ \midrule
\multicolumn{3}{c}{\textbf{Open Source Program Offices (OSPOs)}} \\ \\ \midrule
National government & Supporting the implementation of any national policy for OSS and software reuse. Typically resides with the PSO(s) responsible for digital government and transformation in a country. & France, Iceland \\
Regional or local government & Supporting the implementation of any regional or local policy for OSS and software reuse. Typically resides in the office for IT and/or digitisation efforts. & Denmark, Netherlands \\
Associations of PSOs & Associations  with PSOs as members or owners, enabling these to initiate and collaborate on OSS projects addressing common needs & Denmark, France \\
Institution-centric & Internal departments responsible for IT service provisioning to the overarching institution. Primary goal to build and scale capacity inside the institution in adopting and collaborating on OSS. & Netherlands, Sweden \\ \\ \midrule
\multicolumn{3}{c}{\textbf{Guidelines and support documents}} \\ \\ \midrule
Inbound-focused & Guidelines to consider or follow when acquiring and reusing OSS. & France \\
Outbound-focused & Guidelines to consider or follow when releasing public sector software as OSS. & New Zealand \\ \\ \midrule
\multicolumn{3}{c}{\textbf{Communities of practice}} \\ \\ \midrule
Public sector & Networks focused within the public sector, within or across various levels of government. & Sweden, Netherlands \\
External & External networks, including representatives, e.g., from industry, civil society, and the broader OSS ecosystem. & France \\ \\ \midrule
\multicolumn{3}{c}{\textbf{Software catalogues}} \\ \\ \midrule
Prescriptiveness & Legally mandated vs. voluntary sharing of OSS via catalogues. & Spain vs. Sweden \\
Accessibility & Software shared via catalogue only accessible to PSOs vs. public. & Spain vs. Sweden \\
Software scope & Sharing of public sector software in general vs. OSS only via catalogues. & Spain vs. Sweden \\
Maintenance & Maintenance of the software catalogue by an OSPO or through crowdsourcing. & France vs. Sweden \\
Code repository & Provisioning of code repositories for active OSS development. & Germany \\
Meta-data & Use of meta-data files to describe the software shared enabling decentralised search. & France \\ \bottomrule
\end{tabular}
\end{table*}

\subsubsection{Institutionalised support functions on different levels of government}
In industry, the use of support functions and centres of competency is a well-established practice for implementing a company’s OSS strategy in line with the overarching business goals. These functions are commonly referred to as Open Source Program Offices (OSPOs), a construct and practice that has also transitioned to the public sector and can be found at various levels of government, providing support for the use and release of OSS, and promoting software reuse within government, in line with any overarching government policy~\cite{linaaker2024public}. The different types of OSPOs complement each other in supporting different parts of the government by providing interfaces to each other, sharing resources and knowledge, and more effectively implementing their specific policies and, where applicable, any overarching policy.

The responsibility for supporting the implementation of any national policy for OSS and software reuse typically resides with the PSO(s) responsible for digital government and transformation in a country. These PSOs, or the units within responsible for the support, may be referred to as national-government OSPOs. In France, this is specifically constituted by the Free Software Unit within DINUM~\cite{linaaker2024public}, whereas in other countries the role is more blurred on the organisational level, as with Digital Iceland in Iceland~\cite{icelandgov2021inboundoutbound} and Red.es in Spain~\cite{spain2022guidelines}.

In the Netherlands, an Institution-centric OSPO~\cite{linaaker2024public} has been established within the Ministry for the Interior and Kingdom Relations, with an internal focus on the ministry and its related national-level PSOs. The OSPO, however, is a main driver of implementing the country’s ''Open, unless'' policy and provides implicit support for other parts of the government as well~\cite{dutchgov2022inboundpolicy}. They are, however, also in the process of supporting the establishment of a national government OSPO under the Office of the Government CIO, which would have broader responsibility for implementing the policy. In Sweden, there was no national government OSPO either, although what may be referred to as a series of institution-centric OSPOs exists primarily at the national-level PSOs, such as the Agency for Digital Government, the Swedish National Insurance Agency, and Statistics Sweden.

The lower levels of government also showed that they have OSPOs in place in various cases. In Spain, regional OSPOs were exemplified by the Galician regional government. Larger municipalities were also included in the study, including Barcelona, Amsterdam, Paris, and Aarhus. Local governments, however, seldom have the resources or capabilities to provide the necessary support. Instead, a common approach is to pool their resources and set up association-based OSPOs~\cite{linaaker2023public} where they can share knowledge and initiate, develop, and collaborate on OSS. Examples include ADULLACT in France, OS2 in Denmark, and the Dutch Association of Municipalities (VNG).

\subsubsection{Guidelines for how to interpret and act according to policies }
Several countries maintain guidelines and recommendations for implementing their overarching policies. Outbound policies generally have the most detailed guidelines for the aspects and steps to consider when releasing public sector software as OSS. These guidelines typically have two main parts: one clarifying the legal context and the other supporting the decision on whether a piece of software should be released as OSS. The second part typically focuses more on how to release the software as OSS and on building a community if that is a desired goal for the software.

The former part concerns whether the policy is advisory or prescribes releasing public sector software as OSS. In the Netherlands, the Ministry of the Interior and Kingdom Relations has developed process charts and detailed guidance to support its ''Open, unless'' policy~\cite{dutchgov2022inboundpolicy}. In France, the Free Software Unit provides three criteria related to the software's usability for other OSS projects, the general need for it, and the technical profile of end-users~\cite{dinum2023doc}. Based on the criteria, they propose four levels of openness for how the software may or should be shared.

In the UK, the Government Digital Service maintains a Service Standard that specifies the requirement for public authorities to \textit{``[m]ake new source code open''} in order \textit{``for people to reuse and build on''} the code, notably by publishing the code in an open repository and retaining ownership of the associated intellectual property rights, making it available for re-use under an open license~\cite{ukgovdigitalservices2022standard}. It provides more detailed guidance on implementing this requirement in the Service Manual~\cite{ukgovdigitalservices2017designmanual}.

\subsubsection{Leveraging an external community for knowledge sharing and support provision}
Concerning the practical process for releasing OSS, many guidelines provide rich advice both in themselves, such as in New Zealand~\cite{newzealandgov2016inboundoutbound} and France~\cite{dinum2023doc}, but also by highlighting external sources of best practice, as done by Digitaliseringsstyrelsen in Denmark~\cite{danishgov2022vejledning}. In the former cases, the external ecosystem has been further actively involved in the development of the guidelines. In New Zealand, the guidelines stem from a crowdsourcing process facilitated by an external OSS expert who was brought in for the task. In France, the corresponding guidelines have been iteratively developed and validated by various actors within and outside the government.

An important source of knowledge in the process has been the BlueHats network, a cross-sector community of individuals and organisations focused on the adoption and development of OSS in the public sector~\cite{dinum2021Bluehats}. Related to BlueHats, the Free Software Unit at DINUM also facilitates a Free Software council, bringing together experts and actors from across the public sector and the larger OSS ecosystem. The board's role is to provide advice on topics of concern within the intersection of OSS and digital transformation of the public sector~\cite{dinum2021council}.

The case of Blue Hats exemplifies the importance and value of leveraging an external community to help support the implementation of OSS policies. The NOSAD network in Sweden provides another example of how public servants can interact and share knowledge among themselves and with the larger OSS ecosystem. The network facilitates regular meetups, operates communication channels, and has an online knowledge base with resources to enable reuse and collaboration of OSS and open data. The Netherlands has adopted another network structure for knowledge sharing through its OSPO network, which brings PSOs together with internal OSPOs.

Another example of enabling reuse and collective knowledge sharing is represented through the association-based OSPOs. OS2 in Denmark, for example, has created standardised processes and structures for governance and collaboration on the development of OSS projects. These help both the members (most of which are municipalities) to initiate projects addressing common needs, come together on projects, and engage with suppliers on terms and conditions understood and recognised by both sides in a procurement process. The Dutch Association of Municipalities is on track to establish similar processes and structures based on lessons learned from a pilot project.

Despite the many initiatives and policies supporting software reuse, the cases further show that their sustainability varies with funding. Malta and Iceland, for example, both initiated projects in the early 2010s with the ambition to grow and enable the adoption of OSS and reuse, but both dissipated a few years later. In Iceland, support was continued by Digital Iceland, while in Malta, MITA is not providing any active support. 

\subsubsection{Software catalogues as means for promoting reuse}
Several countries maintain software catalogues covering software developed and/or used by Public Sector Organisations (PSOs). In Spain, the use of the national catalogue is mandated by law, requiring all PSOs to publish acquired applications to enable reuse by other PSOs\footnote{\url{https://administracionelectronica.gob.es/ctt/CTTprincipalEs.htm?urlMagnolia=/pae_Home/pae_SolucionesCTT.html}}. Source code, documentation, license conditions, and associated costs should be shared and declared. The national catalogue is maintained by the Technology Transfer Centre, a state-level PSO. PSOs can also maintain their own versions and integrate with the national catalogue. Several catalogues are maintained by regional governments and are also integrated into the national catalogue.

While the Spanish catalogue is closed to PSOs only and not limited to OSS, the French counterpart, code.gouve.fr\footnote{\url{https://code.gouve.fr}}, is publicly open and explicitly focuses on OSS used and/or developed by French PSOs. The catalogue is maintained by the national government OSPO, which is constituted by the Free Software Unit within DINUM. All OSS listed in the catalogue have adopted the public-code.yml metadata standard\footnote{\url{https://yml.publiccode.tools/}} for public sector OSS projects, which facilitates findability and adoption. By including the metadata file in the catalogue of an OSS project, it can be queried and included in other catalogues, enabling interoperability among regional, national, and international catalogues and further promoting reuse. Other countries, such as Italy and the Netherlands, have also adopted the standard, improving cross-border reuse and adoption of OSS projects.

In Estonia, OSS solutions developed for the government are made public and freely available at koodivaramu.eesti.ee\footnote{\url{koodivaramu.eesti.ee}}. The Estonian government recognises the value of open principles, enabling businesses to adapt these solutions more easily and potentially increasing the export of digital government solutions. Similarly, in the Netherlands, the Developer Overheid\footnote{\url{https://developer.overheid.nl/}} platform provides a library of both APIs and OSS catalogues from various PSOs across the Dutch public sector. There is a long-term goal to evolve the platform into a common source code storage and collaboration platform, possibly based on the OSS social coding platform GitLab. The German government, through their Centre for Digital Sovereignty, has adopted a similar approach with their OpenCode platform\footnote{\url{https://opencode.de/en}}. The European Commission has also created its own environment\footnote{\url{https://code.europa.eu/}}. Currently, however, most public sector OSS projects are hosted on GitHub, as in most cases investigated in this study, although some exceptions use public or internally hosted instances of GitLab.

A less formal but generally recognised example is offentligkod.se\footnote{\url{https://offentligkod.se}} in Sweden, a software catalogue listing OSS used and/or developed by Swedish PSOs. The catalogue was initiated by the Swedish PSO-centred knowledge-sharing network NOSAD. All reports are contributed on a volunteer basis either by the PSOs directly or the vendors providing services based on the OSS. The catalogue is referenced by the Swedish National Procurement Office in its framework for the acquisition of OSS-based software and services.

\begin{framed}\noindent\textbf{Summary}: 
Many policy support initiatives are in place or emerging among the surveyed countries. Some initiatives have been fragile in terms of support and funding, leading to dormancy in some cases, while in others, the support efforts have been picked up in later years. The report identifies the emergence of support functions and centres of competency for OSS and software reuse, also referred to as Open Source Program Offices (OSPOs). These OSPOs have developed at national, institutional, and local government levels, playing a crucial role in building institutional capacity for software reuse through OSS. Association-based OSPOs specifically help less capable PSOs to pool resources and enable sustainable maintenance and governance of common OSS projects.
\end{framed}

\section{Discussion and Suggested Indicators}
\label{sec:discussion}

Below, we discuss our findings in relation to existing literature, highlighting key insights and recommendations emerging from our cross-comparative analysis of OSS policies and support actions. We specifically reflect on how policy design and implementation aspects interact, and how distinctions such as inbound and outbound policies help structure this discussion. Following this, we present a comprehensive set of analytically derived indicators that reflect the main themes of our findings, aiming to provide actionable guidance for governments and a foundation for future benchmarking efforts.

\subsection{Policy motivations and design}

\textbf{Recognise OSS policies as established but uneven in scope and role.} OSS-focused policies are not unique or to be considered an obscurity. Our survey shows that they are widely present internationally at the local, regional, and national levels, aligning with earlier studies~\cite{bouras2014policy, blind2021impact, thevenet2024progress}. Our study further aligns with earlier work~\cite{lostri2023gov} on how policies are formed and passed, e.g., as parliamentary bills, directives and regulations, strategies (general or OSS-specific), or more informal expressions of support for OSS. At the same time, our cross-comparative analysis highlights how these policies vary substantially in form and scope, underscoring the need for and potential of complementary policies to guide government and public entities that may be organizationally and politically disconnected.

\textbf{Make policy direction explicit to better guide design and analysis.} The need for complementary policies is further nuanced due to their inbound and outbound natures, i.e., whether they are focused on the acquisition and procurement of new software, or the release of software developed through public funds. Policies will have different details depending on the direction, with the former typically considering how the procurement process applies or should be tailored, while the latter will focus more on whether, and in what form, software can be released (e.g., for security or privacy reasons). In earlier surveys, the distinction has been more generally labelled under procurement and R\&D~\cite {lostri2023gov}, whereas our findings show the need to make this directional distinction explicit (in line with the open innovation paradigm) to more clearly communicate policy intent and support a more fine-grained understanding of policy design.

\textbf{Balance intervention types with support for effective outcomes.} The type of intervention that the policies come with further varies, ranging from softer high-level enforcements to prescriptive requirements, e.g., on how OSS solutions should be considered in an acquisition process, or whether software developed through public funds should be released as OSS. In line with Van Loon and Toshkov~\cite{van2015adopting}, we observe that softer statements may imply limited to no effect when there is no support for the policy. On the other hand, we know from earlier work that the expected benefits of acquiring OSS solutions or releasing software as OSS can be easily outweighed by potential costs and risks~\cite{linaaker2020share}. Pragmatic case-by-case decisions are the norm in business-driven contexts, which highlights why open-by-default policies should be applied in government, but also underscores the need to balance prescriptiveness with feasible implementation conditions.

\textbf{Align policy rationale with clear decision conditions and goals.} A balance needs to be struck, and especially aligned with the overarching goals and motivations of the policies. In business, any OSS strategy is, in the end, profit-driven and influenced by the strive for market success. For governments in our survey, incentives range from promoting interoperability and growing digital sovereignty and vendor independence to increasing transparency and cost efficiency. Whatever the rationale, the policies need to make clear the decision space and the conditions under which they apply. France and the Netherlands (among many others) require software developed with public funds to be open by default, either automatically or on request. In both cases, conditions are provided for when exemptions can be made and for what reasons, aligning with the overarching policy goals.

\textbf{Extend OSS policy beyond the public sector to the wider ecosystem.} While most of our surveyed policies are focused on the governments and public sector, the cases of South Korea and Japan signal the options of also promoting OSS adoption and development in the national industries. Our findings highlight how many of the policy goals—such as sovereignty, interoperability, and competitiveness—are dependent on a skilful and competitive vendor ecosystem. Small and medium-sized enterprises knowledgeable about OSS development are, for example, critical for competitive, cost-efficient tenders while also ensuring vendor independence for governments and public entities. National policies should, hence, take a more holistic approach to enabling OSS adoption, development, and collaboration across the whole of society. 

\textbf{Integrate procurement and capability building as core policy levers.} Policies should further align with Blind et al.~\cite{blind2021impact}, considering how procurement processes as well as skills development can be promoted and tailored. Our analysis shows that these two dimensions are tightly coupled in practice. Inbound procurement-focused policies can provide guidance and commitment from public entities seeking OSS-based solutions, motivating increased investment by SMEs. Skills strategies, training programs, and calls for national capacity building can help increase the availability of a skilled workforce, required for governments, industries, and vendors to adopt OSS and leverage it per their defined goals.

\subsection{Support actions for policy implementation}

\textbf{Recognise implementation support as essential for policy realisation.}
A policy does not get implemented automatically. Accordingly, the expected outcomes are not realised by default. There is a need for substantial support efforts to comprehend any policy, translate its recommendations into actionable advice, and act on such advice. Our findings reinforce the gap between policy intent and realisation, highlighting the need for implementation support when assessing OSS policy in practice.

\textbf{Institutionalise internal expertise through OSPOs.}
The findings align with earlier work in that internal experts and champions are critical for a sustainable adoption of OSS in public entities~\cite{oram2011promoting, van2015adopting}. Formal support centres, also referred to as Open Source Program Offices (OSPOs), provide a source for such knowledge and advocacy, and by extension constitute an important policy enabler, helping the intended audience for any policy to understand it and implement it in practice~\cite{linaaker2024public}. In this study, we observed OSPOs at both the national and local government levels, providing support for open source within and across public entities, aligning closely with the archetypes proposed by earlier work~\cite{linaaker2024public}. At the same time, our cross-case analysis highlights their role not only as support functions but also as key institutional mechanisms that bridge policy design and implementation.

\textbf{Leverage collaborative steward structures to sustain OSS efforts.}
The role of associations such as OS2 in Denmark~\cite{frey2023we} and iMio in Belgium~\cite{viseur2023communesplone} in supporting its members and owners to collaborate and adopt OSS solutions has also been emphasised in earlier work~\cite{linaaker2024public}. Our findings confirm their role and report on their existence, e.g., in France through ADULLACT and in the Netherlands through VNG. These associations may also serve as a model for collaboration and the adoption of OSS within or across other levels and parts of government. In a sense, they can be compared to OSS foundations (e.g., the Linux or Eclipse foundations), which steward projects long-term, provide neutral governance structures, and enable formal collaboration and resource pooling to support the maintenance of joint OSS projects~\cite{persson2024soft}. While public entities can and should initiate OSS projects, it is typically not within their resources or scope to drive and own them long-term. Having a host-organisation set up for such a prime purpose therefore makes sense on a national and, potentially, also regional (cross-border) level, as in Europe, where the drive towards joint and interoperable solutions is being called for through the Interoperable Europe Act and related efforts to strengthen digital sovereignty.

\textbf{Provide actionable guidance to operationalise policies.}
Serving as centres of excellence and support, the OSPOs (regardless of type) typically provide the processes and tools for both adopting, developing and collaborating on OSS. Among the surveyed cases, we observed guidelines on various levels of detail (and quality), supporting both inbound and outbound OSS policies. For example, when an inbound policy requires OSS solutions to be considered first-hand in an acquisition process, how can such solutions be identified, analysed, and compared with each other and proprietary solutions, considering (among other things) security, license requirements, total cost of ownership, and the need for support and education? Or, the opposite: when an outbound policy requires acquired or internally developed solutions to be released as OSS, on what platform is this done, and under what license? And how will the software stay maintained, and how will contributions from others be reviewed and managed? Our findings highlight that addressing such questions systematically is critical to translating policy into practice.

\textbf{Enable reuse and visibility through shared infrastructure.}
Beyond guidelines, tools and services can also provide an important support for public entities. In this study, which looked in detail at how sharing and reuse could be enabled, software catalogues provided a recurring theme, improving the findability of released solutions. Examples highlight the potential of federated sharing through the public-code.yml standard, which has recently been implemented by the European Commission\footnote{\url{https://interoperable-europe.ec.europa.eu/eu-oss-catalogue}}, providing a critical step toward the development of interoperable solutions, as envisaged for 2030. These observations informed our identification of concrete support mechanisms that can be operationalised and measured as part of OSS enablement.

\textbf{Promote collaboration across sectors to build sustainable capacity.}
Our study further emphasises that public entities cannot act in isolation; they need to collaborate and pool resources, building joint capabilities, as suggested by the joint steward structures and OSPOs across various levels and parts of society. Part of this discourse also includes connections and collaboration with non-public actors, such as industry, academia, and civil society. Among the former, the vendor and service supplier ecosystem is of utmost importance, as it is typically the one developing and maintaining OSS solutions on behalf of public entities~\cite{linaaker2025public}. The long tradition of outsourcing, conservative and risk-averse culture, and lack of sustainable leadership support provide some of the rationale, in alignment with related work~\cite{persson2024soft, cassell2008governments, marco2020outsourcing}. Collectively, these findings highlight the need for coordinated and cross-sector approaches to support the sustained implementation of OSS policies.

\subsection{Indicators for OSS enablement}
With this study, we set out to provide input to global indexes (as the ones we leveraged in our sampling) on how OSS can be leveraged and enabled as a tool for governments and public entities in their digital transformation. Overarching goals include promoting sharing and reuse, and the collaborative adoption and development of joint, transparent, sovereign, and interoperable solutions for their digital public infrastructure and services. Based on the 16 case studies, we synthesise potential indicators reflecting good practice from mature real-world examples, many of which are confirmed and validated through related work. The potential indicators also serve as direct recommendations for practice, which is also a goal for the concerned indices, in providing guidance for their various target groups. 

We further note that the list of indicators is comprehensive and needs to be narrowed and prioritised for any index seeking to assess governments' maturity of OSS adoption in a practical and meaningful way. The exercise of prioritisation is left for practice and future work.

Below, we present potential indicators across 13 areas, divided into two categories: policy incentives and design, and policy implementation and support. Each indicator is defined to be measurable and comparable between countries on an annual basis. 

\subsubsection{Policy incentives and design}


\begin{itemize}
    \item 1. Inbound Policies for Acquiring OSS
    \begin{enumerate}[label=1.\arabic*]
        \item Maturity scale (e.g., 0=None, 1=Drafted, 2=Adopted, 3=Implemented, 4=Monitored for results)
        \item Definition scale (e.g., 0=None, 1=Strategy, 2=Government instruction, 3=Legislative)
        \item Prescription scale (0=No policy, 1=Advisory, 2=OSS with conditions, 3=Mandatory OSS-first, 4=Mandatory with reporting/monitoring)
        \item Coverage breadth (\% of government organizations, \% of regional/municipal agencies covered)
        \item Presence of measurable targets or KPIs related to acquisition and use of OSS in policy documents (Yes/No)
        \item Number and frequency of public progress reports on inbound OSS policy implementation
        \item Frequency of inbound OSS policy review/updates (yearly, every two years, etc.)
    \end{enumerate}
\end{itemize}
    

\begin{itemize}
    \item 2. Outbound Policies for Sharing and Publicising OSS
    \begin{enumerate}[label=2.\arabic*]
        \item Maturity scale (e.g., 0=None, 1=Drafted, 2=Adopted, 3=Implemented, 4=Monitored for results)
        \item Definition scale (e.g., 0=None, 1=Strategy, 2=Government instruction, 3=Legislative)
        \item Prescription scale (0=No policy, 1=Advisory, 2=OSS-first with conditions, 3=Mandatory, 4=Mandatory with reporting/monitoring)
        \item Coverage breadth (\% of government PSOs, \% of regional/municipal PSOs covered)
        \item Percentage of government-developed/procured solutions released as OSS annually (\%)
        \item Cumulative reuse of software solutions between PSOs (annual)
        \item Presence of measurable targets or KPIs related to acquisition and use of OSS in policy documents (Yes/No)
        \item Number and frequency of public progress reports on OSS policy implementation
        \item Frequency of outbound OSS policy review/updates (yearly, every two years, etc.)
    \end{enumerate}
\end{itemize}


\begin{itemize}
    \item 3. Externally-Oriented OSS Policies
    \begin{enumerate}[label=3.\arabic*]
        \item Existence of OSS-targeted provisions in national innovation/industrial policy (Yes/No)
        \item Value/number of grants/subsidies awarded to OSS-related projects (annual)
        \item Industry engagement rate (measured via survey: \% of businesses reporting interaction with government OSS initiatives)
    \end{enumerate}
\end{itemize}

\subsubsection{Policy implementation and support}

\begin{itemize}
    \item 4. Existence of public sector OSPOs (Open Source Program Offices)
    \begin{enumerate}[label=4.\arabic*]
        \item Number of OSPOs formally established at each government level (national, regional, local), and Institutes for Higher Education (annual count)
        \item Percentage of central government domains (e.g., health, justice, education) with an active OSPO
        \item Number PSOs formally supported by these OSPOs per year
        \item Mandate clarity score: Based on review, \% of OSPOs with a documented mandate aligned to policy goals
        \item Coverage ratio: Proportion of all eligible entities (by level) that have an OSPO
    \end{enumerate}
\end{itemize}

\begin{itemize}
    \item 5. Existence of neutral steward bodies for joint OSS projects
    \begin{enumerate}[label=5.\arabic*]
        \item Count of independent legal/administrative bodies acting as OSS stewards (annual) for public sector OSS projects
        \item Number of public sector OSS projects hosted or supported by steward bodies
        \item Stakeholder engagement: Number/type of diverse PSOs participating in joint initiatives coordinated by these stewards
        \item Annual budget or resource allocation for such collaborative public sector OSS projects
    \end{enumerate}
\end{itemize}

\begin{itemize}
    \item 6. Subnational entities leading and scaling OSS development
    \begin{enumerate}[label=6.\arabic*]
        \item Number and share (\%) of municipalities/regions that actively engage in at least one OSS project (yearly, collected via survey)
        \item Annual investment in OSS projects at the subnational level (collected via survey)
        \item Share of subnational entities reporting OSS contributes to (collected via survey):
        \begin{itemize}
            \item Cost savings (\%)
            \item Digital innovation (\%)
            \item Improved interoperability (\%)
            \item Enhanced technical sovereignty (\%)
        \end{itemize}
    \end{enumerate}
\end{itemize}

\begin{itemize}
    \item 7. Support and capacity-building for PSO OSS policies
    \begin{enumerate}[label=7.\arabic*]
        \item Number of support programs or guidance documents issued to PSOs on OSS policy/strategy formulation per year
        \item Proportion of PSOs (national/regional/local) provided with OSS policy support in the past 12 months
        \item Number and percentage of PSOs that established OSPOs following national alignment guidelines
        \item Frequency and participation rate in workshops, webinars, or consultations on OSS policy and reuse
    \end{enumerate}
\end{itemize}


\begin{itemize}
    \item 8. Engagement in national and international OSS ecosystems
    \begin{enumerate}[label=8.\arabic*]
        \item Number of OSPOs/PSOs participating in: national OSS alliances, international OSS organisations, or cross-border OSS networks
        \item Number of participants in international OSS-related events (annual)
        \item Number of collaborative OSS development efforts with international partners per year
        \item Projects or initiatives cited/used by foreign PSOs or partners
    \end{enumerate}
\end{itemize}

\begin{itemize}
    \item 9. Inbound policy guidelines for OSS adoption
    \begin{enumerate}[label=9.\arabic*]
        \item Existence and versioning of guidelines (Yes/No; publication date; most recent update)
        \item Checklist existence: whether a procurement checklist for OSS evaluation is available (Yes/No)
        \item Percentage of procurements referencing guidelines when considering new software
        \item Surveyed clarity and usability of these guidelines for PSOs and developers (1–5 scale)
        \item Coverage of key topics (evaluation, requirements, cost, interoperability, data, etc.—scored in a checklist)
    \end{enumerate}
\end{itemize}

\begin{itemize}
    \item 10. Outbound policy guidelines for public software release as OSS
    \begin{enumerate}[label=10.\arabic*]
        \item Existence and versioning of outbound release guidelines (Yes/No; publication date; most recent update)
        \item Number of PSOs applying the guidelines in releasing software
        \item Licensing and community-building support described in guidelines (Yes/No; checklist coverage)
        \item Surveyed clarity and usability of these guidelines for PSOs and developers (1–5 scale)
    \end{enumerate}
\end{itemize}

\begin{itemize}
    \item 11. Skills Development and Training
    \begin{enumerate}[label=11.\arabic*]
        \item Share of government workforce trained in OSS skills (annual \%)
        \item Number of OSS training sessions/courses delivered to internal/external stakeholders (annual)
        \item Coverage scale (government, industry, academia, civil society – scored separately)
        \item Existence of formal feedback/evaluation mechanism after OSS training (Yes/No)
        \item Reported improvement in OSS-related competencies post-training (survey-based scoring)
    \end{enumerate}
\end{itemize}

\begin{itemize}
    \item 12. Catalogue of public sector OSS
    \begin{enumerate}[label=12.\arabic*]
        \item Existence of a national catalogue for public sector OSS (Yes/No; launch/version date)
        \item Total and annually added number of public sector OSS solutions listed
        \item Catalogue inclusivity score: Assessed on the presence of regional, local, and cross-sector contributions
        \item Compliance of listed projects with Public-code.yml metadata standard (1–5 rating via audit)
        \item Usage rate: \% of PSOs accessing or uploading to the catalogue annually
    \end{enumerate}
\end{itemize}

\begin{itemize}
    \item 13. National social coding and version management platform for government OSS
    \begin{enumerate}[label=13.\arabic*]
        \item Existence and operational status of platform (Yes/No; launch date)
        \item Total and annually added number of OSS projects hosted on the platform
        \item Transparency metrics: Audit logs publicly available, external contributor onboarding time (measured in days)
        \item Accessibility/barrier score: Surveyed ease of access and use for new PSOs or developers (1–5 scale)
    \end{enumerate}
\end{itemize}

\section{Conclusions and future outlook}
\label{sec:conclusions}

The study presents a survey of policies and practices related to software reuse, focusing on OSS in 16 digitally mature countries. The resulting analysis, grounded in desk research and interviews, provides a comprehensive overview with individual case studies for each country.

The survey uncovers an evolving landscape of OSS policy in government and public sector contexts, affirming that OSS-focused policies are neither rare nor marginal, appearing at multiple levels of governance and through diverse channels, from formal legislation to informal endorsements. These policies are inherently directional—addressing both inbound processes (acquisition and procurement) and outbound processes (opening and releasing software)—a distinction we make explicit and use to structure the analysis across cases.

A principal challenge facing policymakers is to clarify both the scope and rationale of OSS initiatives, ensuring that open-by-default principles are balanced with necessary exemptions aligned to transparency, sovereignty, interoperability, and cost-efficiency goals. As some cases demonstrate, achieving these policy aims depends on a mature, skilled vendor ecosystem, further reinforcing the notion that national strategies must take a holistic view across government, industry, academia, and civil society.

Effective implementation of OSS policies further depends on robust support structures and mechanisms internally within governments. The surveyed cases confirm both the importance and the diversity of Open Source Program Offices (OSPOs), as well as the critical roles of administrative and legal bodies, professional associations, and dedicated networks in public sector collaboration. These actors serve as essential enablers for translating policy intention into practice, which we synthesise into a structured framework of policy and support dimensions spanning policy design, institutional actors, and enabling mechanisms.

Sustainable adoption also hinges on specialised training and shared knowledge, facilitating institutional capacity-building and bridging connections with external stakeholders. Findings are synthesised into a comprehensive set of potential indicators for digital government indexes to leverage and guide governments in adopting and enabling OSS as a policy and digital transformation tool. In contrast to existing work and current indexes, which provide limited coverage of OSS, these indicators offer a more comprehensive and operationalised basis for benchmarking and follow-up.

More broadly, the study contributes a cross-comparative and practice-oriented perspective that integrates policy and implementation aspects, providing a foundation for both future research and practical efforts to design, implement, and evaluate OSS-oriented digital government strategies.




\bibliographystyle{elsarticle-num} 
\bibliography{mybibliography}



\end{document}